\documentclass[
  journal=pasa,
  manuscript=research-article, 
  year=2024,
  volume=37,
]{cup-journal}

\usepackage{amsmath}	
\usepackage{amssymb}
\usepackage[caption = false]{subfig}
\usepackage{threeparttable}
\usepackage{comment}

\usepackage{microtype,siunitx,booktabs}
\title{A Fast Radio Burst monitor with a Compact All-Sky Phased Array (CASPA)}

\author{R. Luo}
\affiliation{CSIRO Space and Astronomy, PO Box 76, Epping, NSW 1710, Australia}
\alsoaffiliation{Department of Astronomy, School of Physics and Materials Science, Guangzhou University, Guangzhou 510006, China}
\email[Rui Luo]{rui.luo@gzhu.edu.cn}

\author{R.~D. Ekers}
\affiliation{CSIRO Space and Astronomy, PO Box 76, Epping, NSW 1710, Australia}
\alsoaffiliation{International Centre for Radio Astronomy Research, Curtin University, Bentley, WA 6102, Australia}
\email[Ron Ekers]{ron.ekers@csiro.au}

\author{G. Hobbs}
\affiliation{CSIRO Space and Astronomy, PO Box 76, Epping, NSW 1710, Australia}

\author{A. Dunning}
\affiliation{CSIRO Space and Astronomy, PO Box 76, Epping, NSW 1710, Australia}

\author{C.~W. James}
\affiliation{International Centre for Radio Astronomy Research, Curtin University, Bentley, WA 6102, Australia}

\author{M.~E. Lower}
\affiliation{CSIRO Space and Astronomy, PO Box 76, Epping, NSW 1710, Australia}

\author{V. Gupta}
\affiliation{CSIRO Space and Astronomy, PO Box 76, Epping, NSW 1710, Australia}

\author{A. Zic}
\affiliation{CSIRO Space and Astronomy, PO Box 76, Epping, NSW 1710, Australia}

\author{M. Sokolowski}
\affiliation{International Centre for Radio Astronomy Research, Curtin University, Bentley, WA 6102, Australia}

\author{C. Phillips}
\affiliation{CSIRO Space and Astronomy, PO Box 76, Epping, NSW 1710, Australia}

\author{A.~T. Deller}
\affiliation{Centre for Astrophysics and Supercomputing, Swinburne University of Technology, P.O. Box 218, Hawthorn, VIC 3122, Australia}

\author{L. Staveley-Smith}
\affiliation{International Centre for Radio Astronomy Research, The University of Western Australia, 35 Stirling Highway, Crawley WA 6009, Australia}

\received {dd Mmm YYYY}
\revised  {dd Mmm YYYY}
\accepted {dd Mmm YYYY}
\published{dd Mmm YYYY}


\newcommand{\D}{\mathrm{d}}
\newcommand{\EQ}[1] {Equation~(\ref{#1})}
\newcommand{\FIG}[1] {Figure~\ref{#1}}
\newcommand{\TAB}[1] {Table~\ref{#1}}
\newcommand{\SEC}[1] {Section~\ref{#1}}

\def\deg{^\circ}
\def\gpcyr{\mathrm{Gpc}^{-3}\, \mathrm{yr}^{-1}}
\def\pccm{\mathrm{pc}\,\mathrm{cm}^{-3}}

\def\DM{\mathrm{DM}}

\def\DMe{\DM_{\rm E}}

\def\DMh{\DM_{\rm host}}

\def\fz{f_z}

\def\fw{f_\mathrm{w}}

\begin{document}

\begin{abstract}
Fast Radio Bursts (FRBs) are short-duration radio transients that occur at random times in host galaxies distributed all over the sky. Large field of view instruments can play a critical role in the blind search for rare FRBs. We present a concept for an all-sky FRB monitor using a compact all-sky phased array (CASPA), which can efficiently achieve an extremely large field of view of $\sim10^4$ square degrees. Such a system would allow us to conduct a continuous, blind FRB search covering the entire southern sky. Using the measured FRB luminosity function, we investigate the detection rate for this all-sky phased array and compare the result to a number of other proposed large field-of-view instruments.  We predict a rate of a few FRB detections per week and determine the dispersion measure and redshift distributions of these detectable FRBs. This instrument is optimal for detecting FRBs in the nearby Universe and for extending the high-end of the FRB luminosity function through finding ultraluminous events. Additionally, this instrument can be used to shadow the new gravitational-wave observing runs, detect high-energy events triggered from Galactic magnetars and search for other bright, but currently unknown transient signals. 
\end{abstract}

\section{Introduction}
\label{sec:intro}
In the past decade, the mysterious fast radio bursts (FRBs;  \citealt{Lorimer+07Sci}) have become one of the most fascinating research topics in astronomy \citep{ Thornton+13Sci,Petroff+19A&ARv, CC19ARA&A, Petroff+22A&ARv}. Although these radio flashes last only a few milliseconds (or even tens of microseconds), they can release as much energy as the Sun radiates in a time scale of days to years \citep{Luo+18MN}. There have now been hundreds of FRBs discovered and published, but their origin remains unresolved.  The discovery of a bright radio burst detected from the Galactic magnetar SGR 1935+2154 \citep{CHIME/FRB20Nat, Bochenek+20Nat} provides a clue and further evidence for magnetars as the source of some FRBs. Cosmological FRBs come from various host galaxies \citep{Bhandari+22AJ} and only two active repeaters were found to be associated with persistent radio sources \citep{Marcote+17ApJ, Niu2022}. However, the majority of (extragalactic) FRBs are too distant to detect any multiwavelength counterpart, or to make a detailed study of the source environment. One of the closest, FRB~20200120E, has been revealed to originate in a globular cluster \citep{Kirsten2022_GC}, challenging the magnetar-from-supernovae hypothesis. 

\begin{table*}
    \centering
    \begin{threeparttable}
    \begin{tabular}{lccccccc}
        \hline
        Instrument\tnote{(a)} & CASPA & Parkes CryoPAF & SKA-Low \tnote{(b)} & DSA-110 & GReX & BURSTT-256 & CHIME far-sidelobe \\
        \hline
        Elements & 65 & 98 & 256 & 110 & 1 & 256 & 1024 \\
        Centre Freq.\,(GHz) & 0.9 & 1.35 & 0.2 & 1.4 & 1.35 & 0.55 & 0.6 \\
        Bandwidth (MHz) & 400\tnote{(c)} & 400 & 40 & 187.5 & 1300 & 400 & 400 \\
        $N_{\rm chan}$ & 4096 & 4096 & 512 & 6144 & 16384 & 1024 & 1024 \\
        $t_\mathrm{res}$ (ms) & 0.06 & 0.06 & 10  & 0.03 & 0.01 & 10 & 0.983 \\
        $T_{\rm sys}$ (K) & 25 & 15 & 300\tnote{(d)} & 25 & 25 & 150 & 50 \\
        SEFD (Jy) & 29018 & 26 & 2300 & 111 & $\sim$2M & 5000 & 22500 \\
        $N_{\rm pol}$ & 2 & 2 & 2 & 2 & 2 & 1 & 2 \\
        $N_{\rm beam}$ & 72 & 72 & 3600 & -- & 1 & -- & -- \\
        FoV (deg$^2$) & 10368 & 2 & 11909 & 10.6 & $\sim$20000 & $\sim$10000 & 1800 \\
        \hline
    \end{tabular}
    \end{threeparttable}
    \caption{A comparison of the key system specifications used in the simulations of "all-sky" transient monitors}
    \begin{tablenotes}
    \footnotesize
    \item (a) Some instruments listed are not "all-sky", such as Parkes CryoPAF, DSA-110 and CHIME far-sidelobe. We include them here to provide a comparison with some higher sensitivity but narrower FoV instruments.
    \item (b) The key parameters are adopted from \cite{SPW22aapr}.
    \item (c) The maximum possible bandwidth is larger, but the sky is not fully sampled at the top of the band so only 400MHz is used in the simulation.
    \item (d) At this frequency, the system temperature is set by the diffuse cosmic radiation and will vary significantly with sky position and frequency. 
    \end{tablenotes}
    \label{tab:inst}
\end{table*}

The efficiency of blind transient searches have been enhanced by upgrades to existing instruments and the development of widefield facilities, such as, the 13-beam receiver of Parkes radio telescope \citep{Staveley-Smith+96PASA} which detected the first FRB (\citealt{Lorimer+07Sci}) to the phased array feed for the Australian Square Kilometre Array Pathfinder (ASKAP, \citealt{Hotan2021ASKAP}). 
At present, the FRB sample size is expanding rapidly, mainly thanks to the Canadian Hydrogen Intensity Mapping Experiment (CHIME), which has discovered by far the most FRB sources to date. CHIME consists of four cylindrical parabolic reflectors, each with a 256-element linear array which provide its large field of view ($\mathrm{FoV}\sim200\, \mathrm{deg}^2$, \citealt{CHIME/FRB21ApJS}). 
The forthcoming mega facilities, i.e., the Square Kilometre Array (SKA), may be able to monitor and detect even larger numbers of FRBs, providing the wide FoVs search options that are implemented at full sensitivity. \cite{Sokolowski+21PASA} explores this option using an SKA-Low station. Except all-sky instruments described above, some other instruments with large FoVs are being constructed for surveys of the transient universe. The Deep Synoptic Array 2000 (DSA-2000), which consists of thousands of 4.5-metre dishes, is a facility to carry out radio camera survey with fast scan speed \citep{Hallinan+19BAAS}. The DSA-110, as a pilot version for DSA-2000 currently, has been demonstrated to be a powerful instrument to discover dozens of FRBs with precise localisation \citep{Law+24ApJ} and good polarization measurements \citep{Sherman+24ApJ}.

The impact of a telescope's FoV and the on-sky observing time is different for surveys of transient (one-off or sporadic) events compared to persistent sources \citep{Cordes07AAS}. A survey for sporadic events never ends and the number of events or sources (e.g., FRBs) found is proportional to the product of the observing time and the FoV. In contrast, discovering a new persistent source in a given survey area is only possible with increased sensitivity and that only improves as the square root of the observing time or equivalently the square root of the FoV for a given search area.
Hence the value of FoV as a discovery space strategy for sporadic events is much more important than it is for surveys of persistent sources such as active galactic nuclei (AGN) or even pulsars. Hence, in the telescope design, the trade-off between FoV and sensitivity will be different and all-sky, all-the-time monitor is more competitive for some scientific objectives than much higher sensitivity telescopes with smaller FoV.

An all-sky monitor can be constructed using a radio array formed by small antennas. \cite{1995AcAau..35..745D} described such an omni-directional radio telescope, the \textsc{Argus} telescope, and reported on successful observations using eight narrow bandwidth elements. At the time, however, the processing requirements were prohibitive for a larger array with broader bandwidth. More recently, a few all-sky transient instruments with $\sim 10^4 \, \mathrm{deg}^2$ are being planned. For instance, the single element Galactic Radio Explorer (GReX) is designed to find the brightest bursts in our local Universe \citep{Connor+21PASP}, such as  the Galactic FRB detected by STARE2 \citep{Bochenek+20Nat}, and potential extremely luminous FRBs in nearby galaxies. Another all-sky facility, the Bustling Universe Radio Survey Telescope for Taiwan (BURSTT), is proposed to detect and localise hundreds of bright FRBs per year \citep{Lin+22PASP}. 
Recently, \cite{Lin+23arXiv} reported ten new FRBs discovered in the far sidelobes of CHIME. In this case each of the four CHIME line feeds alone act as a 256 element, one-dimensional, all-sky monitor. 

A possible technology for an ``all-sky'' monitor could be based on 
the Cryogenically-cooled Phased Array Feed (CryoPAF) that is now being commissioned as a focal plane array for the Parkes 64-m radio telescope (``Murriyang'').  At the focal point of the telescope, the CryoPAF provides a relatively small FoV (although much larger than a single pixel receiver). But if situated on the ground looking up, it could be used to monitor a large fraction of the sky. In this case, the array could be significantly enhanced, since it would not be constrained to illuminate a fixed size dish with no spill-over and with dimensions limited to space at the focus of the telescope. The performance and science cases for such a compact all-sky phased array (CASPA) is the focus of this paper.

The basic properties of some proposed all-sky instruments are listed in Table~\ref{tab:inst}. We include “all-sky monitors” in this table and use the specifications in our simulations. There are many other FRB survey instruments (CHIME, FAST, ASKAP, MeerKAT, DSA...) with higher sensitivity in a much smaller FoV, but we have only included the Parkes CryoPAF and DSA-110 to illustrate the very different parameter space being probed by these instruments. The Parkes CryoPAF provides a very convenient comparison since it has identical frequency coverage, backend and processing requirements as CASPA. It is well beyond the scope of this paper to include all other FRB search instruments.
The specifications in Table 1 are used in our FRB detection simulations. Note that some of these specifications are simplified from a real system (see Table~\ref{tab:caspa}) since it is hard to simulate the complex frequency behavior of the sensitivity and FoV for a wide bandwidth system.

The structure of the manuscript is organised as follows. We describe the specification of an optimised beam forming phased array on the ground in \SEC{sec:as-paf}. In \SEC{sec:frbdet}, we perform the Monte Carlo simulations on detectable FRBs for this ground-based phased array, and for some other proposed ``all-sky'' monitors. We discuss the localisation of FRBs in \SEC{sec:localisation}. We discuss the broader range of science cases for such an instrument in \SEC{sec:science} and we summarise the impact and future outlook for such an instrument in \SEC{sec:conclude}. 

\section{A compact all-sky transient monitor using phased array technology}
\label{sec:as-paf}

We do not include a detailed design study for an all-sky monitor, but instead provide a  baseline representation of a realisable system based on the technology already developed for the Parkes CryoPAF \citep{Dunning23}. The Parkes CryoPAF has a close-packed regular grid of antenna elements with 196 ports, 98 for each polarization. It generates 72 focal plane array beams and the digital backend will implement FRB and pulsar search modes for all 72 beams.  In contrast, our proposed receiver will be uncooled but without the focus area constraints, it can have a larger diameter and significantly improved performance compared to the CryoPAF (see \TAB{tab:caspa}).

We will minimise the number of beams required to cover the FoV, in order to reduce the beam forming and processing requirements, which are often the limiting factor for radio telescope performance. This will require the most compact array possible as long as the receiving elements remain nearly independent at all frequencies. 

The diameter of this array, $D$, gives the width of the beams in the Zenith direction of
\begin{equation}
\Theta_{\rm Z} = \frac{\lambda }{D \cos \theta_{\rm z}}\,,
\end{equation}
where $\theta_{\rm z}$ is the angle from the zenith. Accordingly, the beam width in the Azimuthal direction is
\begin{equation}
\Theta_{\rm A} = \frac{\lambda }{D}\,.
\end{equation}
In order to calculate how many beams are required to cover the large FoV, we use a coordinate transform so that the beam area is independent of the sky position. In this coordinate system where a unit sphere on the sky is projected down to a unit circle on an X-Y plane, the phased array beams will be circular and independent of $\theta$.  In this projection area of sky seen by each beam ($A_\mathrm{beam}$) is
\begin{equation}
    A_\mathrm{beam}=\frac{\pi \lambda^2}{4 D^2}.
\end{equation}
The total FoV (as an area in the unit circle projection plane) measured from the zenith down to a zenith angle $\phi_\mathrm{FoV}$ is
\begin{equation}
 A_\mathrm{FoV}=\pi\sin^2\phi_\mathrm{FoV}.
\end{equation}
Thus, for a given FoV, the number of beams required is
\begin{equation}
 N=\frac{A_\mathrm{FoV}}{A_\mathrm{beam}} \approx \frac{4D^2 \sin^2 \phi_\mathrm{FoV}}{\lambda^2}.
\end{equation}
If we know the required FoV, the number of beams we can process and the observing frequency, then we can work backwards to obtain the diameter of the array $D$ and its area $A_{\rm arr} = \pi D^2/4$.


For optimum sensitivity, these beams must be independent so that the number of independent receiving elements ($N_{\rm ele}$) should equal the number of beams ($N_\mathrm{beam}$).  In practice, this will be reduced by the array packing efficiency. The hexagonal packing efficiency for circles, $\eta = \pi/2 \sqrt{3} = 0.91$, so we can only fit $\eta N_\mathrm{beam}$ elements into the circular area of diameter $D$. Given the number of elements and the system temperature then we can obtain the sensitivity of the system noting that the system equivalent flux density (SEFD) is estimated using traditional single dish formula\footnote{It should be noted that the traditional interpretation of a single dish SEFD at the beam centre will be different for a multiple beam phased array with its relatively flat sensitivity across the FoV at the low frequency end of the band but varying sensitivity across the FoV at the higher frequencies.}.

Since we have already developed a backend for the Parkes CryoPAF which implements FRB search mode on 72 beams, we set $N_\mathrm{beam} = 72$ in this case. The field of view we propose to cover is 25\% of the sky ($\phi_\mathrm{FoV} = 60$ deg) and the observing frequency for optimum sensitivity is near the low-end of the observing band (0.75\,GHz). The equations above therefore lead us to an array extent of $D = 2.0$\,m, giving an approximate effective array area of $A_{\rm arr} \simeq 3 $\,m$^2$. The number of receiver elements assuming hexagonal close packing is $N_{\rm ele} =  65$. Adopting a system temperature of 25\,K would give a SEFD of $\sim 25000$\,Jy for this phased-array system (see \TAB{tab:caspa}). 

As defined above, this SEFD will scale as 1/$f_c^2$ if the number of elements, beams and sky coverage remains constant\footnote{This scaling relation assumes the temperature of receiver is frequency independent and the temperature conversion follows the Rayleigh-Jeans inverse square frequency dependence.}. If we critically sample at the high frequency, then the effective area remains constant with frequency, but this is inefficient because we will be oversampled at the low frequency. 
The simulation parameters that we use later (and listed in the left-most column of Table~\ref{tab:inst}) have therefore been restricted to the lower part of the available bandwidth. In \TAB{tab:caspa}, we list the parameters of a realistic array, but note that this is not the detailed modelling that would be required for a final system design. In particular, the frequency range and bandwidth specified in \TAB{tab:caspa} are based on the CryoPAF receiver array which is already being commissioned, but the final CASPA system will more likely be optimised for a slightly lower frequency.



\begin{table}
    \centering
    \begin{tabular}{cc}
    \hline
        System & Specification \\
        \hline
        Frequency & 0.7 -- 1.4 GHz \\ 
        Elements & 65  \\
        Polarization & 2 \\
        Bandwidth & 700 MHz \\
        $t_\mathrm{res}$ & 0.06 ms \\
        $N_{\rm chan}$ & 4096 \\
        $T_{\rm sys}$  & 25 K \\
        Filling factor & 91\%  \\
        Diameter & 2.0 m\\
        $A_\mathrm{arr}$ (hexagon) & $\sim$3.0 m$^2$ \\
        Beamwidth (0.7 GHz) & 14 deg \\
        Beamwidth (1.4 GHz) & 7 deg \\
        $N_{\rm beam}$ & 72 \\
        FoV & 10368 deg$^2$ \\
        Fraction of sky & 25\% \\
        SEFD (0.75\,GHz) & 25000\,Jy \\
        rms sensitivity in 1 msec & 40\,Jy\\
        \hline
    \end{tabular}
    \caption{The specifications of the Compact All-Sky Phased Array - CASPA}
    \label{tab:caspa}
\end{table}

\subsection{FRB searching with phased array beams}

Instead of computing images 
 from correlation measurements of the coherence function across the aperture every integration cycle, we propose to use a fixed set of real-time beamformers.  These could be digital, taking advantage of the fixed regular array to use FFT techniques, or even analogue using wide bandwidth time delay beamformers. 
The time resolution ($t_\mathrm{res}$) for the FRB DM search is therefore not limited by image processing speed and can be optimised for the expected FRB pulse widths. For the other more sparse arrays listed in \TAB{tab:inst} which require realtime image computation to coherently combine visibilities, the highest time resolution achievable may be significantly longer than some of the FRB pulse widths and this will decrease the detection signal to noise.  Quoted minimum integration times are included in the table and range from 10\,$\mu$s for GReX to 10\,ms for BURSTT-256 and the SKA-Low station.
 The time resolution of 0.06\,ms given in \TAB{tab:inst} and \TAB{tab:caspa} and used in the simulation is the value for the Parkes CryoPAF beamforming backend.

\subsection{Radio Frequency Interference}

A wide-band, all-sky monitor will be open to radio frequency interference (RFI) coming from any direction, but as already emphasised by \cite{1995AcAau..35..745D} the planar array on the ground has many advantages.  It has low gain towards the horizon when situated on the ground reducing the effect of terrestrial interference. Tests with the Parkes CryoPAF have confirmed that the RFI environment improved when the system was on the ground compared to being up at the focus cabin of the 64m dish. Satellite and airborne interference will still be a major problem, but  the direction and characteristics of the RFI signal are immediately known because one beam will always be pointing towards the RFI signal.  This can provides powerful RFI mitigation using anti-coincidence logic or adaptive filtering techniques.




Any residual RFI in a single station could still generate false triggers, but as discussed in Section~\ref{sec:localisation}, multi-station arrays will be able to confirm any detections from extraterrestrial signals and such a geographically dispersed instrument will be essentially immune to false detections due to RFI.

\section{FRB Detectability}
\label{sec:frbdet}

\begin{figure}
    \centering
    \includegraphics[width=1\textwidth]{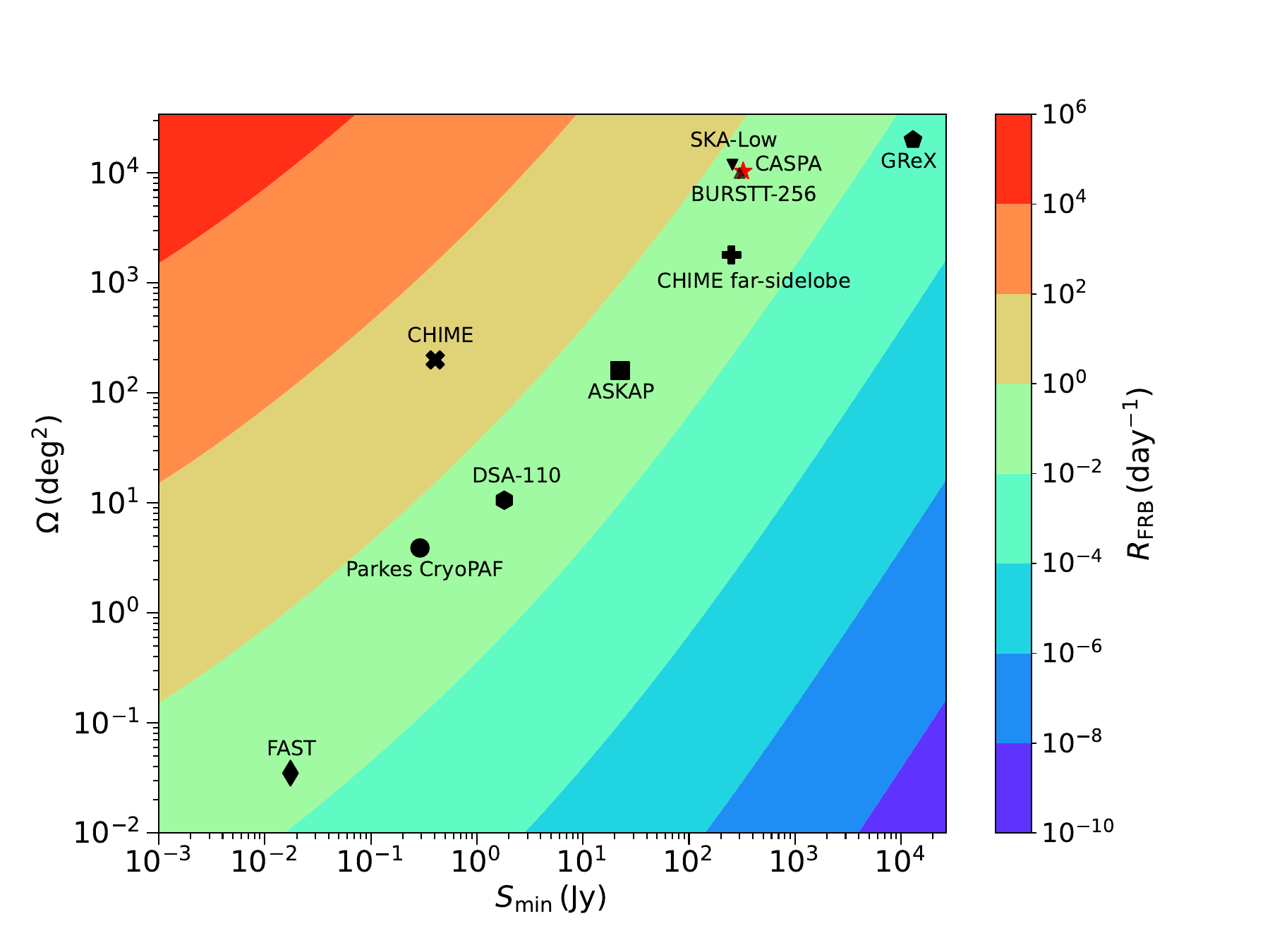}
    \caption{The detection rate contour of instruments. The x-axis represents the flux threshold in units of Jansky, the y-axis is the FoV in units of square degree, and the colour bar denotes the inferred detection number per day. The detection rate of several instruments are marked, such as the CASPA (star in red), Parkes CryoPAF (circle), GReX (pentagon), BURSTT-256 (triangle up), SKA-Low (triangle down), DSA-110 (hexagon), CHIME far-sidelobe (plus), CHIME (cross), ASKAP (square) and FAST (diamond).}
    \label{fig:dr_map}
\end{figure}

 Here we consider the properties of the FRBs that will be detected with our ground-based phased array and compare with predictions for the other instruments listed in \TAB{tab:inst}. In \FIG{fig:dr_map} we show the FoV and the limiting flux density for the six systems listed in \TAB{tab:inst} along with the CHIME, FAST and ASKAP telescopes. We also separately show the properties of the primary beam for the CHIME telescope as well as the system which accounts for the far side-lobes. The  detection rates inferred by integrating the FRB luminosity function \citep{Luo+20MN} are given in \TAB{tab:det_rate}. 

\begin{table}
     \centering
     \begin{threeparttable}
     \begin{tabular}{c|c}
         \hline
         Instrument & $R (\mathrm{day}^{-1})$\tnote{a} \\
         \hline
         CASPA & $0.34^{+0.38}_{-0.21}$ \\
         Parkes CryoPAF & $0.19^{+0.22}_{-0.12}$ \\
         SKA-Low & $0.54^{+0.60}_{-0.33}$  \\
         DSA-110 & $0.16^{+0.18}_{-0.10}$ \\
         GReX & $0.004^{+0.004}_{-0.002}$ \\
         BURSTT-256 & $0.47^{+0.52}_{-0.29}$ \\
         CHIME far-sidelobe & $0.08^{+0.09}_{-0.05}$ \\ 
         \hline
     \end{tabular}
     \begin{tablenotes}
     \footnotesize
     \item (a) The error bars (68\% confidence) are dominated by the propagation of errors from the measured event rate density of FRB luminosity function in \cite{Luo+20MN}.
    \end{tablenotes}
    \end{threeparttable}
    \caption{The predicted FRB detection rates of the instruments listed in \TAB{tab:inst}.}
     \label{tab:det_rate}
 \end{table}
 
 



\subsection{The Monte Carlo Simulations}
\label{sec:simu}


In order to obtain the properties of the detectable FRBs for a given system, we implement the following recipe:

\begin{enumerate}[label=(\roman*)]
    \item Sample the FRB luminosities, $L$, according to the Schechter function as follow, \\
    \begin{equation}
    \phi(L)\,\D L=\phi^* 
    \left(\frac{L}{L^*}\right)^{\alpha}e^{-\frac{L}{L^*}}\,\D \left(\frac{L}{L^*}\right)\,,
    \label{eq:lf}
    \end{equation}
    where $\phi^*=339_{-313}^{+1074}\,\gpcyr$, $\alpha=-1.79_{-0.35}^{+0.31}$ and $\log L^*=44.46_{-0.38}^{+0.71}$ according to \cite{Luo+20MN}.
    \item Sample the intrinsic FRB pulse widths in the local rest frame of FRBs using the log-normal distribution constrained in \cite{Luo+20MN}: \\
    \begin{equation}
	\fw(\log w_\mathrm{i}) = \frac{1}{\sqrt{2\pi\sigma_{\rm w}^2}}\exp\left[-\frac{(\log w_\mathrm{i}-\mu_{\rm w})^2}{2\sigma_{\rm w}^2}\right]\,,
	\label{eq:wd}
	\end{equation}
	where the measured dimensionless mean value is $\mu_w=0.13_{-0.13}^{+0.11}$ and the standard deviation is $\sigma_w=0.33_{-0.06}^{+0.09}$ \citep{Luo+20MN}.
    \item Consider the cosmological principle for galaxy distribution and possible cosmological evolution for FRB population summarised in \cite{Zhang+21MN}, by sampling the FRB redshifts, $z$. The redshift distribution is given as 
    \begin{equation}
    \begin{aligned}
	\fz(z) &= \frac{\D N}{\D t\D V}\frac{\D t}{\D t_\mathrm{obs}}\frac{\D V}{\D z} \\
	&=\left[(1+z)^{a\eta}+\left(\frac{1+z}{B}\right)^{b\eta}+\left(\frac{1+z}{C}\right)^{c\eta}\right]^{1/\eta} \\
	& \quad \frac{1}{1+z}\cdot\frac{c\,D_{c}^2(z)}{H_0E(z)}\,,
	\end{aligned}
	\label{eq:fz}
	\end{equation}
	where $a=3.4$, $b=-0.3$, $c=-3.5$, $B\simeq5000$, $C\simeq9$ and $\eta=-10$ according to \cite{Zhang+21MN}. We then calculate the DM values corresponding to the contribution from the intergalactic medium (IGM) at the sampled redshifts.
    \item Use the DM distributions of host galaxies at redshift bin $z$ described by \cite{Luo+18MN} to sample the DM values contributed by host galaxies in the local rest frame of the sources.
    We assume that the DM distribution of host galaxies in the nearby Universe is given as a logarithmic double Gaussian function.
	\begin{eqnarray} 
	&&f_\mathrm{host}(\DMh|z=0) = \\ \nonumber
 &&\sum_{i=1}^{2} a_{i}\exp\left\{-\left[\frac{\log (\DMh|z=0)-b_i}{c_i}\right]^2\right\}\,, 
	\label{eq:dmh0}
	\end{eqnarray}
    where $a_1=0.0049$, $b_1=0.8665$, $c_1=1.009$, $a_2=0.0126$, $b_2=1.069$, $c_2=0.5069$ as given for the galaxy case of ALGs(NE2001) in \cite{Luo+18MN}.
    \item Sample the DM values caused locally by the FRB progenitors using the uniform distribution from 0 to 50\,$\pccm$, as assumed in \cite{Luo+18MN}.
    \item Produce Galactic DM values using the YMW16 model \citep{YMW17ApJ}, and then sum the DMs from all of components mentioned above to obtain the total observed values.
    \item Obtain the beam responses by generating a random uniform distribution of FRB positions. For the fixed horizontal arrays we add a factor of $\cos{\theta}$ to compensate for the change of effective collecting area with zenith angle, $\theta$. For the CHIME far-sidelobe monitor, the beam shape is modelled using the results from \cite{Amiri+22ApJ}. 
    \item Compute the received peak flux density using the simulated luminosities, redshifts, and the beam responses of FRB positions within the beam size. Note that we assume a flat spectrum of FRBs (spectral index as 0) here.
    \item Based on the intrinsic pulse widths, redshifts and DMs of FRBs obtained in the steps above, calculate the observed pulse width impacted by DM smearing and scattering broadening. In particular, the DM smearing is given as
    \begin{equation}
        \tau_\mathrm{DM}=8.3 \mu\mathrm{ s}\,\frac{\Delta f_\mathrm{ch}}{\mathrm{MHz}}\,\frac{\mathrm{DM}}{\pccm}\,\left(\frac{f_c}{\mathrm{GHz}}\right)^{-3}\,,
    \end{equation}
    and we adopt the scattering-DM empirical relation from \cite{Krishnakumar+15ApJ} as follows.
    \begin{equation}
        \tau_\mathrm{sc}=3.6\times 10^{-6}\,\mathrm{ms}\,\mathrm{DM}^{2.2}(1+1.94\times 10^{-3}\,\mathrm{DM}^{2.0})\,.
    \end{equation}
    \item Select the FRBs where the peak fluxes are above the instrumental threshold. The threshold of peak flux density is calculated using the radiometer equation as below.
    \begin{equation}
        S_{\mathrm{min}}=\frac{\rm S/N_0\, SEFD}{\sqrt{N_\mathrm{pol}\,\mathrm{BW} w} }\cdot {\rm MAX}\left(1,\sqrt{\frac{t_\mathrm{res}}{w}}\right)\,,
        \label{eq:rdm}
    \end{equation}
    where $\mathrm{S/N}_0$ is the threshold of signal-to-noise ratio, e.g., $\mathrm{S/N}_0=10$ is adopted in this paper, $\mathrm{BW}$ is the bandwidth, $N_\mathrm{pol}$ the number of combined polarization channels, $\mathrm{SEFD}$ is system equivalent flux density and $w$ is the observed width of the FRB.
    For systems with poor time resolution ($t_\mathrm{res}$), such as BURSTT-256 and SKA-Low, the fluence threshold is converted using $t_\mathrm{res}$ as the integration time of the system. 
    \item Generate waiting times of adjacent events during blind search. particularly, the distribution of waiting times follows the Poisson process as below
    \begin{equation}
    f_\mathrm{t}(\Delta t)=\lambda e^{-\lambda \Delta t}.
    \label{eq:poi}
    \end{equation}
    The expected number of events is given as $\lambda=\rho \Omega t$, where $\Omega$ is the FoV in units of deg$^2$ and $t$ is the observing time. The mean event rate $\rho$ is calculated by integrating the luminosity function in units of volumetric rate along redshift bins, that is,
    \begin{equation}
	\rho = \int_0^{\infty} \frac{1}{1+z}\frac{D(z)^2}{H(z)}\D z  \int_{\log L_{\mathrm{min}}}^{\infty} \phi(\log L)\,\D\log L\,.
    \end{equation}
    Note that \begin{equation}
	L_\mathrm{min}(S_{\mathrm{min}}, z)=4\pi D_{\rm L}^2(z) \Delta\nu_0 S_\mathrm{min}\,,\end{equation}
    where the threshold of flux density $S_{\mathrm{min}}$ is described in Step (x). Note that this does not consider any frequency dependence under the assumption of a flat spectrum for FRBs, thus no k-correction is needed in this case. 
\end{enumerate}

\subsection{Detection rate distributions}

\begin{figure}[htbp]
    \centering
    \includegraphics[width=1\textwidth]{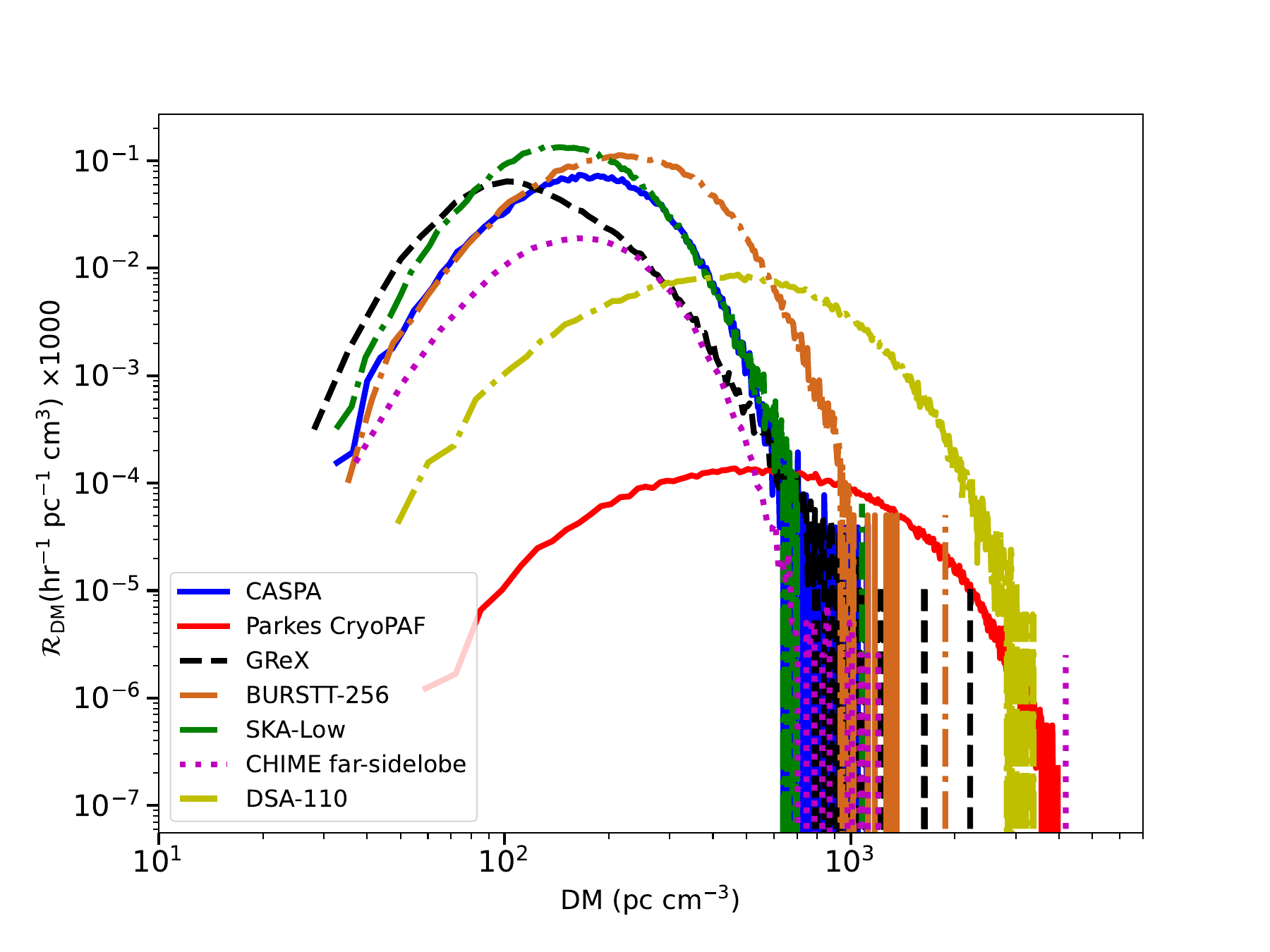}
    \caption{DM distributions of simulated FRB detected by several instruments. The x-axis is the total DM in units of $\pccm$, the y-axis denotes the event rate density in units of per hour per unit of DM.}
    \label{fig:dm_rate}
\end{figure}

We simulate 100,000 FRBs using the Monte Carlo recipe above and then
obtain the detection rate densities of multiple instruments in DM space, which are shown in \FIG{fig:dm_rate}. Note that the event rate density of the DM distribution is calculated using 
\begin{equation}
    \mathcal{R}_\mathrm{DM} = P(\mathrm{DM})\cdot\frac{N_\mathrm{FRB}}{t_\mathrm{obs}}\,,
\end{equation}
where $P(\mathrm{DM})$ is the probability density of DM distribution function, given by $\int P(\mathrm{DM})\mathrm{d}\,\mathrm{DM}=1$. $N_\mathrm{FRB}$ and $t_\mathrm{obs}$ are the total number of simulated FRBs and total observing time in the simulations, respectively. The expected average detection rate of specific instrument in \TAB{tab:inst} is obtained by integrating the curves in \FIG{fig:dm_rate}.  

The peak of detection rate density for each instrument can reflect the integrated detection rate directly, for instance, the SKA-Low and BURSTT-256 systems have the highest peaks in \FIG{fig:dm_rate} and the highest predicted detection rates from \TAB{tab:det_rate}. Although the event rate of a instrument is determined by both FoV and sensitivity, the range of DM distribution is almost dominated by sensitivity. Our simulations show the DM distribution for the Parkes CryoPAF ranges from hundreds to thousands of $\pccm$ with a peak around $800\,\pccm$, which is consistent with previous Parkes detections \citep{Arucs+22MNRAS}. By contrast, for all-sky monitors such as the ground-based phased array or a dipole array, the detectable FRBs are more likely to be low-DM. As highlighted by \FIG{fig:dm_rate}, the Parkes CryoPAF and our proposed ground-based all-sky monitor will be complementary in science cases regarding the very different DM distribution of the detectable FRBs.

We also note that the DM range of BURSTT-256 is wider than that of CASPA with a slightly shifted peak value. The sensitivity of radio instrument is basically determined by both SEFD and time resolution according to the radiometer equation given in \EQ{eq:rdm}. In this scenario, a poorer time resolution can be compensated by a higher SEFD for BURSTT-256. That's why its rate-DM distribution looks close and even a bit better than CASPA from \FIG{fig:dm_rate}. However, given the extreme computational requirements, it will be very hard to achieve this kind of balance. The much lower filling factor in the BURSTT-256 array design will result in much higher computational load. Our simulations here can merely present the results without considering the practical complexity in instrumentation and computation.


\section{Localisation}\label{sec:localisation}

The FRBs detected using the all-sky monitor will be relatively close, as shown in \FIG{fig:dm_rate}.  If the events can also be localised for these nearby FRBs then multi-wavelength observations of the FRB hosts and studies of the progenitor environment will be much more effective. As described in the next section, this all-sky monitor will also allow the electromagnetic follow-up (and hence localisation) of gravitational wave events. We consider some localisation options below.



\subsection{One phased-array station}
 
 At zenith, the phased array beam will have a half power beam width (HPBW) of 7 degrees at 1.4\,GHz \footnote{Although the search mode will be undersampled at 1.4\,GHz, we can reprocess the voltage buffers dumped after a detection with full sampling at any frequency}.  The position of an event within the beam can be determined from the amplitudes in adjacent beams to an accuracy of HPBW/signal to noise. A 10-sigma event will be positioned to an accuracy of 40'.  This will only be sufficient to identify extremely close-by FRB hosts, but it will be more than adequate to search for coincidences with gravitational wave events.  
 
 Since the aperture is fully sampled by the proposed array there is no positional ambiguity due to multiple sidelobes.  Both the position within the beam which detects the FRB and/or gravitational wave event and its fluence will be well determined for all candidates. 

\subsection{Three phased-array stations}

To obtain higher precision, we will need multiple spatially separated stations.  We then have two possible procedures.  We could either use intensity based pulse time of arrival (ToA) measurements or interferometric voltage cross-correlations between stations.   Intensity based ToAs are what e.g. GReX is planning. For FRBs the ToA can be measured to a precision of about 0.1 msec so even with stations separated by 1000s of km this would only provide a localisation precision of about 1 degree which is no better than the single coherent station.  

However wide bandwidth voltage cross correlations will be able to measure delays to better than a wavelength making sub-arcsecond precision position measurement possible with baselines of only 10s of km.  These individual all-sky monitor stations will have insufficient sensitivity for the normal astrometric calibration procedures using astronomical sources, so it would be necessary to tie them to an existing connected element array with a common clock.   An obvious opportunity would be to locate the monitor stations with the outer antennas of the ASKAP array.  While increasing the baseline length to VLBI scales allows increasing localisation precision, maintaining diffraction-limited accuracy would pose an increasing calibration challenge.

Since the transient events will be from point sources and would almost certainly be the only transient in the beam at a given time, three stations are sufficient to determine a 2D position.  
To simplify the processing we envisage a full FRB dispersion measure search being done on all 72 beams at one station (the primary station).  This is preferably the station with the lowest RFI environment.  The other two stations will have simple voltage buffers a few seconds long on each receiver port.  Voltage dumps will be triggered by the primary station and the beam forming and post processing will be carried out off-line.  This greatly reduces the backend cost of the two secondary stations and greatly reduces the data rate to an easily manageable level.  

\section{Discussion on the Science Cases}\label{sec:science}

Given the extremely large FoV, but relatively low sensitivity, the ground-based phased array CASPA would be used for different science cases than the more traditional radio facilities such as CHIME, ASKAP, Parkes, MeerKAT and FAST.  Here we provide a summary of some of the likely science cases.

\subsection{Uncovering FRBs in the nearby Universe}
\label{sec:nearby}

Using the detection rate distributions described in \SEC{sec:frbdet}, we see how the sensitivity of a given system influences the DM range of the FRBs that will be detected.  The all-sky monitors necessarily have relatively low sensitivity and hence a larger number of FRBs with low-DM in the nearby Universe are likely to be discovered. 

\begin{figure}
    \centering
    \includegraphics[width=1\textwidth]{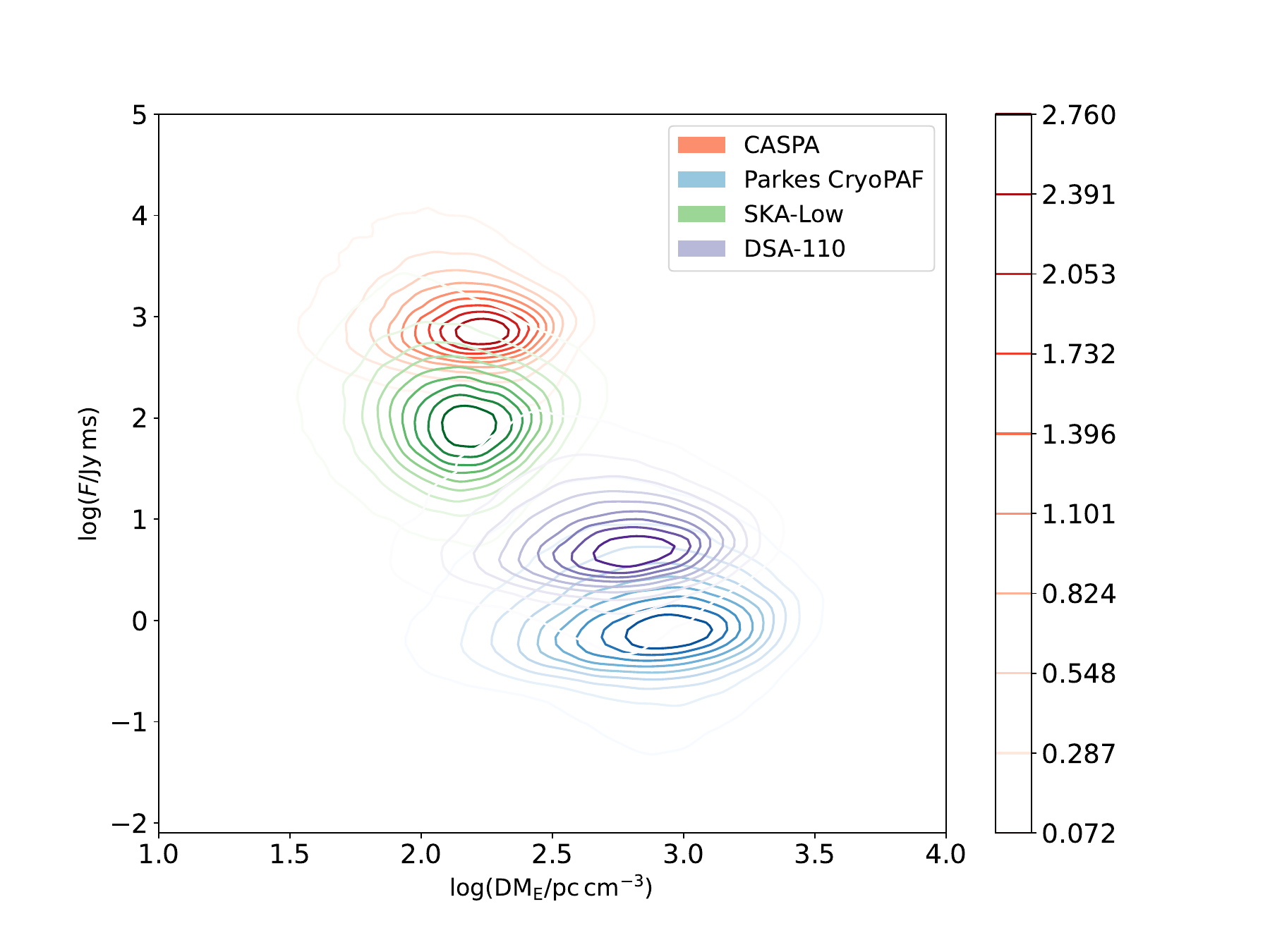}
    \caption{Fluence - $\DMe$ distribution of simulated FRB samples in the logarithmic space. The x-axis represents the extragalactic DM with units of $\pccm$, and the y-axis is the fluence of FRBs in units of Jy\,ms. All the simulated FRBs for CASPA (red), Parkes CryoPAF (blue), SKA-Low (green) and DSA-110 (purple) are clustered as contours with each colour listed in the upper right legend. On the right is the colourbar denoting the estimated kernel density of the FRB sample of CASPA specifically.} 
    \label{fig:fdm}
\end{figure}

To explore the population that CASPA would uncover in more detail, we re-analysed the simulated FRBs for CASPA in the parameter space of fluence versus extragalactic DM, and then compare it Parkes CryoPAF, SKA-Low and DSA-110 (see \FIG{fig:fdm}). Clearly, the Parkes CryoPAF and DSA-110 are likely to detect more high-DM FRBs, which is helpful to study the FRB evolution at high redshift. In contrast, the FRBs detectable for CASPA and SKA-Low have rather low extragalactic $\DMe$ ranging from 30 to 400 $\pccm$, but high fluences from $10$ to $10^4$ Jy\,ms. Such bright bursts with low DMs are mostly originated from the nearby Universe. 

At the time of writing, there have been more than 50 FRBs pinpointed at host galaxies \citep{Gordon+23ApJ} with redshifts up to 1.01 \citep{Ryder+23Sci}. The localised FRB samples at low-redshift ($z<0.1$) are so limited that the population of nearby FRBs is not well characterised. Hence, understanding the properties of these nearby sources is essential to bridge the energy gap between Galactic and cosmological FRBs, and it is also needed for a comprehensive view on the evolution of FRBs. Since the luminosity function that we used in these simulations is constructed from the sample of more distant FRBs, our population modelling is the most conservative case for such FRBs.
Our modelling assumes a smooth volumetric FRB rate, but the star-formation rate in the local volume ($<10$\,Mpc) is higher than a large comoving volume by a factor of 2 \citep{Mattila+12ApJ},  so we may expect to detect even more FRBs from our local Universe and their spatial distribution will not be uniform.

\subsection{Extending the FRB luminosity function}
\label{sec:exlf}

\begin{figure}
    \centering
    \includegraphics[width=1\textwidth]{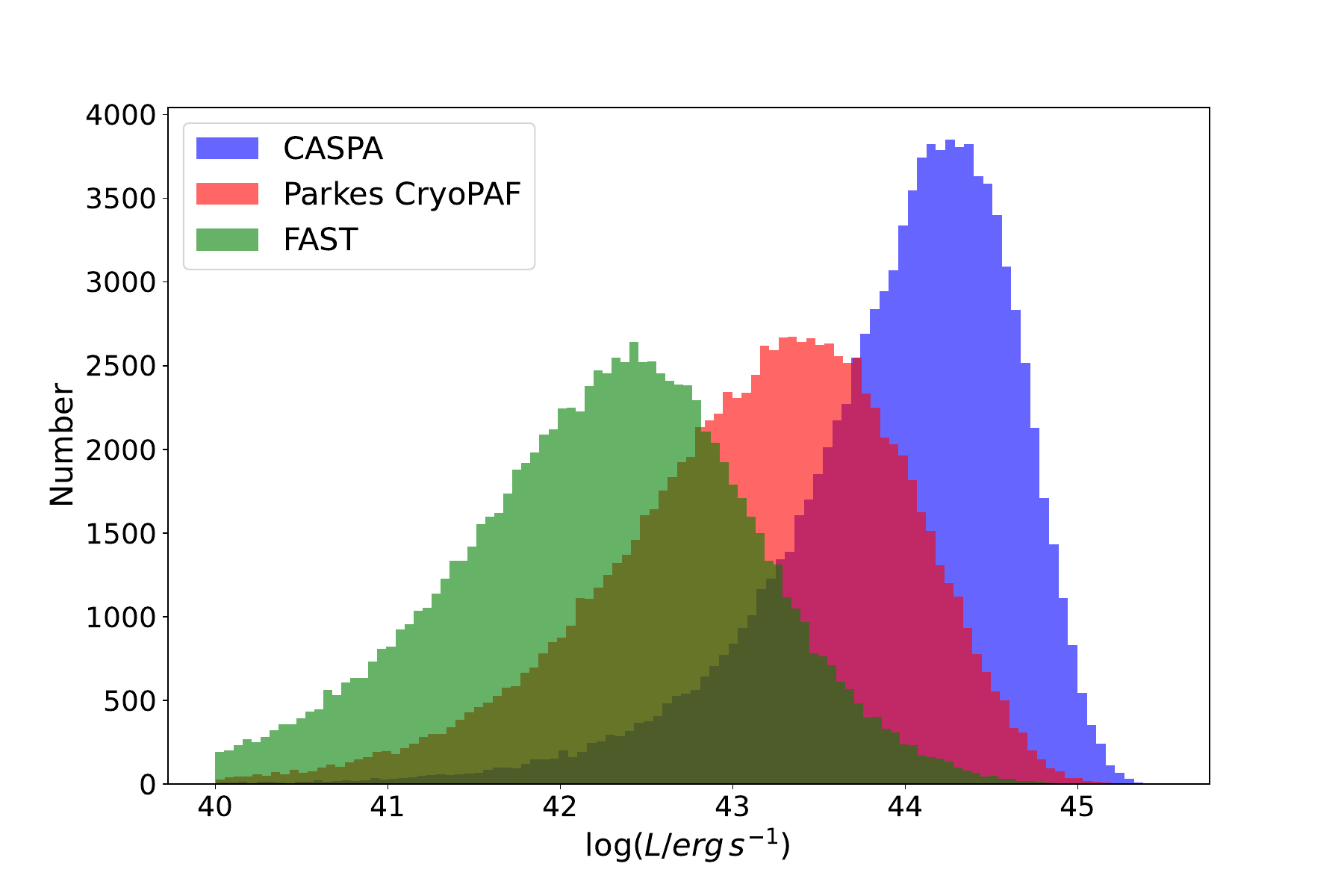}
    \caption{Luminosity distributions of the simulated FRBs detectable for CASPA (blue), Parkes CryoPAF (red) and FAST (green), respectively. The x-axis represents the luminosity of FRBs in logarithmic scale, the y-axis is the number of simulated FRBs detected.}
    \label{fig:lum}
\end{figure}

For FRBs at larger distances, we will only be able to detect ultra-luminous FRBs. Any such ultra-luminous events must be rare requiring a large FoV monitor to find them. We compare the luminosity distributions of three different systems: CASPA, Parkes CryoPAF and the Five-hundred-metre Aperture Spherical radio Telescope (FAST, \citealt{Nan+11IJMPD}) in \FIG{fig:lum}.
The peak of the luminosity distribution for CASPA is close to the higher cut-off of the input luminosity function we used in the Monte Carlo simulations described in earlier. 
This distribution is strongly skewed to the rare highest luminosity FRBs for the less sensitive instruments so they will set the strongest constraints on the high luminosity cut-off. Some studies of the cut-off luminosity from the various FRB samples have been made, e.g., using the ASKAP localised FRBs combined with the Parkes non-localised ones \citep{James+22MNRAS} and using the first CHIME/FRB Catalogue \citep{Shin+23ApJ}. However, the intrinsic cut-off luminosity is not well determined because of selection biases that occur, especially when conducting surveys with the large telescopes. 
A dedicated all-sky monitor such as CASPA would be a powerful instrument to constrain the high-energy limit of the FRB emission mechanism. 

\subsection{Shadowing GW events}
\label{sec:gw}

\begin{figure}
    \centering
    \includegraphics[width=\textwidth]{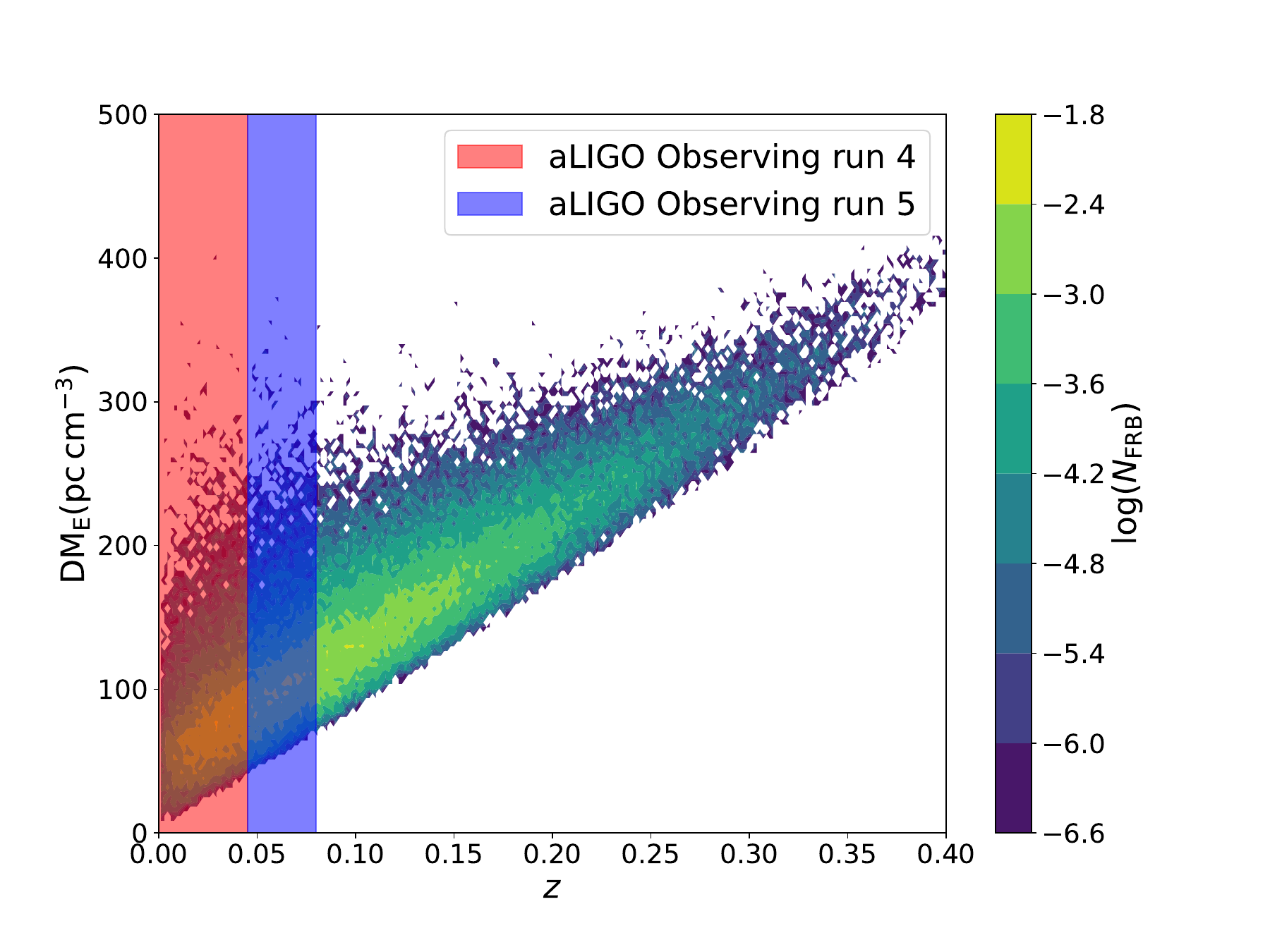}
    \caption{$z-\DMe$ distribution of the simulated FRBs for CASPA. The x-axis denotes the cosmological redshifts and the y-axis denotes the extragalactic DM in units of $\pccm$, and the colour bar is the logarithmic number density of this 2-D histogram. The shaded regions from left to right represent the redshift ranges of aLIGO O4 (red) and O5 (blue) respectively.}
    \label{fig:z-DM}
\end{figure}

From the simulations, we can also obtain the dispersion-redshift distribution for the All-sky Phased Array, which is shown in \FIG{fig:z-DM}. The sample tells us the redshifts of FRBs that would be detected by CASPA will usually be low, peaking at $z=0.06$ with a range from 0 to 0.3. 

The large FoV allows the all-sky monitor to shadow gravitational wave detections by the advanced Laser Interferometer Gravitational-Wave Observatory (aLIGO). The All-Sky monitor bias to detections of nearby events is also an advantage. Adopting the distance limits estimated for aLIGO Observing run 4 (O4) and 5 (O5) in \cite{Abbott+2020LRR}, we have included these distance limits in \FIG{fig:z-DM}. The all-sky monitor can fully cover all the possible FRB-GW association events. There are some theoretical models that account for FRBs as being double neutron star mergers \citep{Totani13PASJ, Yamasaki+18PASJ}. In such a scenario, we would expect to observe possible FRB-GW associated events by both radio telescopes and GW detectors.


Radio counterparts associated with gravitational wave (GW) events involving at least one neutron star or white dwarf have been predicted well before the discovery of FRBs \citep[e.g.][]{2001MNRAS.322..695H}, and scenarios have been proposed to produce emission during the inspiral phase, at point of merger, from the post-merger remnant, and/or from the remnant's subsequent collapse \citep[for reviews, see][]{2016MNRAS.459..121C,2019MNRAS.489.3316R}. However, the sensitivity limit of the current gravitational wave detector network to such mergers is less than 200\,Mpc \citep{2021arXiv211103606T}, meaning that if such events are associated with the known population of FRBs (as suggested by \citealt{Moroianu2023}), their GW signatures will be undetectable.

This suggests that the optimum way to search for radio emission associated with GW events is ``shadowing'' --- constantly monitoring the same sky viewed by the LIGO--VIRGO--KARGA (LVK) network. Our proposed system will be ideal for such a purpose, and we expect to have time-coincident radio data for a large fraction of all GW detections. The large positional errors characteristic of GW detections will be readily covered by the large FoV of this ground-based array. Furthermore, there will be no need to re-point upon receiving a trigger: the instrument will continue to monitor the visible part of the GW localisation region as it passes overhead. This will help overcome cases where public alert information is delayed, as was the case for GW~190425 \citep{2020ApJ...892L...3A}.

If a fraction of the observed FRB population does originate from compact object mergers, their fluence at Earth, if emitted from within the LVK horizon, would be readily detectable by our proposed system according to \FIG{fig:z-DM}. However, FRB-like emission may have difficulty escaping the merger ejecta \citep{2023arXiv230600948B}. In such a scenario, any visible bursts must be produced either pre-merger, or be delayed by perhaps years post-merger. It is impossible for targeted follow-up programs to be sensitive to either scenario \citep{2019MNRAS.489L..75J,Dobie2019}; only an all-sky monitor therefore stands a chance of detecting such radio bursts.

\subsection{Monitoring magnetar flares and burst storms}
Giant flares from Galactic (and possibly extra-galactic) magnetars have been observed at X-ray and gamma-ray wavelengths \citep{1999Natur.397...41H, 2005Natur.434.1098H, 2021Natur.589..211S}.
The short duration (milliseconds to seconds) of the prompt emission from these events, combined with their low event rate makes conducting contemporaneous radio observations extremely difficult with the limited FoV of traditional telescopes.
Non-detections of a coincident radio burst from the 2004 giant flare of SGR 1806$-$20 in the far sidelobe of the Parkes Multibeam set a fluence upper-limit of 1.1-110 MJy\,ms, depending on the assumed attenuation factor \citep{2016ApJ...827...59T}. 
More recently CHIME/FRB reported no detections of a burst coincident with GRB 231115A, suggested to be a giant flare from a magnetar located in M82, down to a limiting fluence of 720\,Jy\,ms \citep{2023ATel16341....1C}.
There has however been some success in performing follow-up observations of magnetars undergoing `burst storms' events where hundreds to thousands of hard X-ray bursts are emitted over the course of a few days.  
Both the April 2020 FRB-like burst and more recent intermediate intensity radio bursts from SGR~1935$+$2154 have been associated with bright X-ray bursts that were emitted during such burst storms \citep{Giri+23arXiv}. 
This proposed all-sky monitor may provide similar radio detections as was the case for the enormously energetic FRB-like burst from magnetar SGR~1935$+$2154 \citep{CHIME/FRB20Nat, Bochenek+20Nat}. 
Notably, this flare was only 40 times less energetic than the weakest extragalactic FRB known at the time. 
If a significant fraction of the extragalactic FRB population follows the same emission mechanism that was involved in the SGR~1935$+$2154, then finding additional events in our galaxy will provide invaluable clues about the progenitors and the emission mechanism of FRBs.

An all-sky monitor situated in the Southern Hemisphere will, for the first time, continuously monitor the entire Southern galactic plane and Magellanic Clouds. 
This would allow for the Galactic event rate and energy distribution to be determined for Galactic magnetars going two orders of magnitude fainter than SGR 1935+2154.



\subsection{Finding the unknown}

Historically, astronomical serendipitous discoveries have always followed any extension of the observing parameter space \citep{KB23sndr.book}. With unprecedented FoV, CASPA will have the potential to 
explore a large parameter space which has not been accessible before and hence would have the potential to find something totally unknown . In recent years, anomalous detections have been reported by the Australian Square Kilometre Array Pathfinder (ASKAP) widefield surveys, e.g., the Odd Radio Circles (ORCs, \citealt{Norris+21PASA}) from the Evolutionary Map of the Universe Pilot Survey (EMU) and a weird polarized radio source \citep{Wang+21ApJ} from the Variables and Slow Transients (VAST).


All-sky monitors are only practical for arrays with small diameter and low angular resolution. Such arrays are completely confusion limited, but short period transients such as FRBs are easily detectable as signal differences on time scales short compared to the motion of the sky through the fixed pattern of beams.  Hence the primary science case described in this paper is to detect FRBs.  However, we may be able to extend this to longer-duration and longer-period transient sources by taking advantage of the fixed pattern of beams and the low angular resolution. The sky will move through the 15-deg beams at the survey frequency in an hour so we could extend the search for transients to much longer time scales. Any strong rare events with time scales similar such as those due to the long duration transients discovered by the Murchison Widefield Array \citep{HWN+22Nat, HWN+23Nat} would be detectable. We could even form a reference baseline as the sky moves through the beams to extend the detection of any unexpected changes to even longer time scales.  The detectability of a range of short-duration events including the unknown has been modelled by \cite{Luo+22MN} and tested by \cite{Yong+22MN}.





\section{Summary and Outlook}\label{sec:conclude}

Large FoV instruments can play a critical role in the blind search both for rare and for nearby FRBs, but it is physically difficult to combine a large FoV with the large apertures needed for high sensitivity.
One solution is a compact phased array on the ground looking up and forming enough independent beams from the coherent combination of all elements to provide the large FoV while maintaining the sensitivity of the total aperture.  We have argued that the optimum configuration for an all-sky monitor is a close packed array with element separation $d=\lambda/2$.  We described such a phased array with 72 active receiver elements working in the frequency range of 0.7-1.4 GHz. This will have a fully sampled extremely large instantaneous FoV of $\sim10^4$ square degrees. By coherently combining all elements, the sensitivity in each of the 72 beams is the same as having a 3 m$^2$ aperture with no additional image processing required. As technology improves, arrays with thousands, or even tens of thousands of elements, corresponding to apertures up to 20-m diameter will become possible.  The FRB dispersion measure search still has to be done at the full 700\,MHz bandwidth in each of the 72 dual polarization beams, hence it is important to minimise the computational requirements without compromising either the dispersion measure search range or the sampling time. We have included an analysis of a representative array configuration, CASPA, which maximises sky coverage with the minimum number of independent signal paths to process. We do not explore design details any further in this paper but the beam forming and processing systems for CASPA have already been developed for the Parkes CryoPAF.


If a similar system is deployed in the Northern Hemisphere, 24 hour observations will cover the entire sky every day, and may detect 4 or 5 FRBs per week. These all-sky monitors will be optimal for detecting bright FRBs in the nearby Universe and for constraining the high end of the FRB luminosity function.  The use of three monitors would allow sub arc-second level localisation of the FRB events allowing multiwavelength follow-up. The unprecedented instantaneous FoV at radio wavelengths opens up a very large parameter space for serendipitous discoveries of the unknown, including short duration techno-signatures. See chapter 6 discussing the Omni-directional SETI Search in \cite{Ekers+2002SETI} and \cite{SPW22aapr}.

\begin{acknowledgement}
R.L. is supported by the National Natural Science Foundation of China (Grant No. 12303042). C.W.J.\ acknowledges support by the Australian Government through the Australian Research Council's Discovery Projects funding scheme (project DP210102103).
\end{acknowledgement}

\bibliography{refs}

\appendix


\end{document}